\documentclass{aa}  

\usepackage{graphicx}
\usepackage{amsmath}
\usepackage{amssymb}
\usepackage{xcolor}
\usepackage{hyperref}

\hypersetup{
    citecolor  = blue,
    colorlinks = true,
    }
\usepackage{xspace}
\usepackage{bm}
\usepackage{txfonts}
\newcommand{\orcidlink}[1]{\protect\href{https://orcid.org/#1}{\protect\includegraphics[width=8pt]{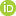}}}

\newcommand{\cs}{\ensuremath{c_\text{s}}}
\newcommand{\co}{\ensuremath{c_0}}

\newcommand{\idefix}{\texttt{Idefix}\xspace}
\newcommand{\kokkos}{\texttt{Kokkos}\xspace}

\renewcommand{\Im}{\ensuremath{\text{Im}}\xspace}

\begin{document}

   \title{Formation of spirals in early stage protoplanetary discs}

   \author{M. Van den Bossche\,\orcidlink{0000-0003-4755-9875}
          \inst{1} \& 
          O. Gressel\,\orcidlink{0000-0002-5398-9225}
          \inst{1}
          }

   \institute{\inst{1} Leibniz-Institut für Astrophysik Potsdam (AIP), An der Sternwarte 16, D-14482, Potsdam, Germany\\              \email{mbossche@aip.de}        		}

   \date{Received XXXX; accepted 23/03/2026}

  \abstract
   {Class II protoplanetary discs feature numerous non-axisymmetric substructures like spirals and the underlying mechanisms for their formation are still highly debated. Coincidentally, early stage, massive discs are subject to the gravitational instability that causes them to collapse into denser substructures. However, like for most instabilities, real systems usually remain marginally stable, here with Toomre parameter $Q \gtrsim 1$.}
   {We study how the self-gravity of the gas triggers the growth of spiral structures in the disc. We specifically focus on discs that are considered stable, that is, with respect to the gravitational instability (with $Q > 1$), as these discs remain unstable to non-axisymmetric perturbations like spirals.}
   {After a linear stability analysis, we produce high-resolution 2D shearing sheet simulations with the GPU-accelerated code \idefix of self-gravitating discs. We probe different initial densities and thermodynamical models of Toomre-stable discs.}
   {The initial transient growth of the spiral wave matches the linear theory provided we take into account the time dependency of the amplification. The spirals are then rapidly non-linearly amplified with growth rate $\approx 10$ orbital time scale. After this time spiral large scale mode are amplified up to 1000 times more than linear theory predicts. At later times, low density discs reach a weak gravito-turbulent state with $\alpha\approx 10^{-3}$ and discs with higher density undergo runaway collapse of the spiral arms. All discs exhibit dominant large-scale spirals.}
   {}

   \keywords{planet formation --
                gravitational instability -- spiral shocks -- accretion
               }

   \maketitle
%

\section{Introduction}

Class II discs show ubiquitous substructures that have been observed with different telescopes for more than 25 years.  More specifically, spirals in protoplanetary systems have first been detected in the Hubble Space Telescope era \citep{grady_hst_1999,clampin_hst_2003}. Then ground-based adaptive optics telescopes such as Subaru observed the same and new systems with increased accuracy \citep{fukagawa_spiral_2004,muto_discovery_2012,grady_spiral_2013}. In the last decade, they were also observed in the infrared with VLT/SPHERE probing ever smaller radii close to the host star \citep{benisty_asymmetric_2015,wagner_discovery_2015}, but also with ALMA \citep{dong_eccentric_2018,huang_disk_2018}. Their presence in infrared observations suggests that not only the gas, but also the dust of these young discs features these spiral structures (see \citealt{bae_structured_2023} for a review).

Yet the physical origin of these apparently ubiquitous spirals remains debated between two main scenarios, that both could lead to spiral growth, but that individually fail to explain the spirals of certain systems.

The first requires the presence of a perturber, that is, a massive enough planet or companion star whose tidal torque excites spiral waves \citep{ogilvie_wake_2002,bae_planet_2018}. If the presence of a external perturber is still a viable explanation for some systems like MWC 758 \citep{wagner_jwst_2024}, recent works showed that for the observed spirals, the corresponding planet should be massive enough to be detected in several systems and were not (see, for example, \cite{dong_observational_2015,dong_dwarf_2016,zhu_structure_2015,meru_origin_2017,uyama_SEEDS_2017}). These systems then require another mechanism to explain the emergence of the observed spiral structures.

The other main mechanism known to produce spiral-waves is the self-gravity of the disc. This traditionally translates to the gravitational instability (GI -- \citealt{gammie_nonlinear_2001}). The criterion for this instability to trigger is the Toomre parameter \citep{toomre_gravitational_1964}.
\begin{equation}
    Q = \frac{\kappa \cs}{\pi G \Sigma  }< 1\Leftrightarrow \text{unstable}\,,
\end{equation}
where $\kappa$ is the epicyclic frequency (equal to the Keplerian frequency $\Omega_\text{K}$ for a Keplerian disc), $\cs$ is the sound speed of the gas, $\Sigma$ is the surface density of the gas, and $G$ is the gravitational constant. The instability is triggered when $Q<1$, that is, when the disc is cold (low $\cs$) or massive enough (high $\Sigma$) such that the pressure can no longer counterbalance the self-gravitational attraction of a perturbed, over-dense region within a disc.

However, at first glance, this mechanism fails to explain the formation of spirals as observations \citep{cleeves_coupled_2016,perez_spiral_2016} suggest that these discs are not massive enough for this instability to trigger.

Yet, numerical and analytical works have shown that spiral waves could still be excited through the self-gravity channel in "stable" discs \citep{durisen_GI_2007,deng_standing_2022}, that is a lower-mass disc than traditionally required for the standard GI with $Q > 1$. This is possible as the initial Toomre stability analysis relies on the assumption of an axisymmetric disc, and, as it turns out, when the azimuthal dimension is re-introduced, the instability arises allowing for spirals to grow in lower density regimes than initially thought possible.

The growth and amplification of these spirals is not a recent discovery, as \cite{goldreich_spiral_1965} already knew of that swing amplification mechanism. Yet, this growth phase has not been revisited with modern tools and most work focused on the long-term properties of self-gravitating flows such as angular momentum transport by the spiral shocks or fragmentation of the disc \citep{gammie_nonlinear_2001,lodato_testing_2004,mejia_thermal_2005,paardekooper_numerical_2012}. The growth phase of these spiral density waves is a compulsory step for a self-gravitating  disc to become turbulent and possibly fragment, and needs better understanding and precise quantification.

In this work, we focus on the initial rise of these non-axisymmetric spiral features of the disc density distribution in ``stable'' discs with $Q > 1$. First, we emphasize that special treatment needs to be included to improve the usual WKBJ analysis in order to capture the initial linear transient growth. Then, we show that this transient linear growth actually give rise to a fully non-linear growth regime that remains absent from past and current theoretical works. We measure and quantify this growth for different disc density and thermodynamics, that can not be included in linear analysis.

This paper is laid out as follows. In the second section, we recall the theoretical framework relevant to studying local properties of thin discs and the linear equation for spiral wave evolution. In the third section, we present the numerical methods used to go beyond the linear approximation. The fourth section presents the results of this study, which are discussed in section~\ref{sec:Discussion}.

\section{Theoretical Framework}

\subsection{Self-gravitating discs}

In this work, we study the local properties of a self-gravitating disc flow. The Hill approximation provides us with the relevant approximation of the disc properties to study the small-scale dynamics associated with spiral shocks. In the local Hill approximation, the dynamical evolution of a non-magnetised cold self-gravitating disc is described by the Euler and Poisson equations

\begin{align}
        \partial_t \Sigma  + \nabla \cdot (\rho \bm{v}) &= 0\,, \label{eq:Mass}\\
        \partial_t \bm{v}  + \bm{v} \cdot \nabla \bm{v} &= - \frac{1}{\rho}\nabla P - \nabla \left(\Phi_\text{Hill}  +\Phi_\text{SG}\right)\,, \label{eq:Momentum}\\
        \Delta \Phi_\text{SG} &= 4 \pi G \Sigma \label{eq:Poisson}\,.
\end{align}

These are for a razor thin disc, that is, $\rho(x,y,z) = \Sigma(x,y) \delta(z)$, where we have vertically integrated all quantities. The Hill potential is $\Phi_\text{Hill} = q \Omega^2 x^2$, where $q$ is the local shear rate and $\Omega$ the local Keplerian frequency.

We note that because the shearing sheet is a periodic geometry (a compact manifold -- with no boundaries), the mean density has to be subtracted from the density field when solving the Poisson equation.

The system is closed by a ideal gas equation of state with different adiabatic index $\Gamma$, as well as isothermal ($\Gamma = 1$). These two approximations allow us to probe different regimes. The isothermal is well-suited to describe a disc where any viscous or shock heating is being compensated for by losses, typically radiative losses. The adiabatic approximation allows for some heating; here we expect spiral shock heating to be the dominant heating term. Adjusting the value of the adiabatic index, $\Gamma$, allows for artificially changing the efficiency of the heating to probe different regimes. 

When the disc is not isothermal, we also solve the energy conservation equation.

\begin{equation}
    \partial_t E + \nabla \cdot \left( E + P\right ) = -\Sigma \bm{v}\cdot \nabla\left(\Phi_\text{Hill}  +\Phi_\text{SG}\right)\, ,
    \label{eq:Energy}
\end{equation}
where $E$ is the total energy density.

\subsection{Linear shearing-wave equation}

Following \cite{gammie_linear_1996,paardekooper_numerical_2012} we derive the linear evolution equation for a spiral wave mode of the form

\begin{equation}
    \Sigma(x,y,t) = \hat{\Sigma}(t) e^{i[k_x(t)x + k_y y]} = \hat{\Sigma}(t) e^{i[(k_{x,0}  + q \Omega k_y t )x + k_y y]} \,,
    \label{eq:SWbasis}
\end{equation}

with $k_{x,0}$ and $k_y$ the components of the initial wave vector of this spiral-wave mode. We note that because of the local shear $k_x(t)= k_{x,0}  + q \Omega k_y t $ is a function of time. In the case of a finite-size shear-periodic domain, it is convenient to refer to these vectors by the associated wave-numbers: $\ell$ and $m$ defined as $k_{x,0}= 2\pi \ell /L_x$ and $k_{y}= 2\pi m /{L_y}$. $L_x$ and $L_y$ are the domain size in each direction.

We obtain the linear equation, for an isothermal disc

\begin{align}
    	&\ddot{\sigma} - 2q \Omega \frac{k_x k_y}{k^2} \dot{\sigma} + \left[  k^2 \cs^2 - 2\pi G \Sigma_0 k + \kappa^2\left ( 1-q \frac{k_y^2}{k^2}\right)\right ] \sigma\nonumber\\
        &+ 2 \Omega \left(1-q \frac{k_y^2}{k^2} \right) \Sigma_0 \xi_1 = 0\,,
        \label{eq:linSW}
\end{align}

where $\sigma = \hat{\Sigma}/\Sigma_0$, with $\Sigma_0$ the density of the background uniform density sheet and coincides with the $k_x = k_y = 0$ mode, and $\xi_1$ is the perturbed potential vorticity, in our case, it is a conserved quantity and we have $\xi_1= -(2-q)\,\Omega \sigma / \Sigma_0$.   We moreover use $k^2 = k_x^2 + k_y^2$, and introduce $\kappa$ the epicyclic frequency, defined such that $\kappa^2 = 2(2-q)\,\Omega^2$. We also use this linear equation to benchmark the self-gravity solver of the code, as detailed in Appendix~\ref{app:PoissonSolver}.

From this amplified-oscillator-like differential equation, we understand that the amplification of the spiral wave comes from the shear $q \neq 0$. We note that non-axisymmetric modes ($k_y \neq 0$) will be amplified (or damped) by the shear.

We note that $\sigma^\Gamma$ does not have an expansion close to $\sigma = 0$ in the general case. This means that the non-isothermal case can not be captured by this linearised equation.

\section{Numerical Methods}

We use the GPU-accelerated finite-volume \idefix code by \cite{lesur_idefix_2023} to solve equations~(\ref{eq:Mass})--(\ref{eq:Poisson}) and (\ref{eq:Energy}).

\subsection{Grid and boundaries}

In this work, we use a uniform Cartesian grid of size $x/H \in [-16 , 16]$ and $y/H \in [-16 , 16]$, where $H = \cs / \Omega$ is the vertical scaleheight of the disc. We thus have a domain of size of $L_x = L_y=32$, and we use $2048 \times 2048$ points to resolve this space. This choice yields square cells with 64 cell points per scaleheight, $H$. As detailed below, we expect from the linear theory that large-scale structures dominate; this resolution and box size allows us to capture them and much smaller scales too. As shown in App. \ref{app:BC}, this resolution is sufficient to capture modes of size smaller than the scaleheight.

We use the shearing-box formalism \citep{hawley_local_1995} in its 2D version (no vertical dimension). The $x$-direction boundaries are shear-periodic with shear rate $q$ and the $y$-direction boundaries are periodic. For this work, we implement the shear-periodic boundary condition for the self-gravity potential. The test of this boundary condition is presented in App. \ref{app:BC}.

In our units, we fix the sound speed $\cs = 1$, and local orbital frequency $\Omega=1$ to be unity. The mass unit is chosen such that the gravitational constant $G=1/\pi$, such that there is a simple relationship between the average sheet density and the Toomre parameter: $\Sigma_0 = Q^{-1}$.

\subsection{Initial conditions}

We use a uniform initial density $\Sigma = \Sigma_0$, corresponding to a uniformly sheared flow $v_y = -q \Omega x$.

We use a polytropic equation of state as the initial condition

\begin{equation}
    P = A \Sigma^{\Gamma}\,,
    \label{eq:EoS}
\end{equation}

where $A$ is a constant which can be expressed with respect to a reference sound speed  $A = \co^2/\Gamma$. The isothermal case reduces to the case $\Gamma = 1$.

We then add a random noise to the velocity fields with an amplitude of 10\% of the sound speed. We additionally use a 10\% noise in the density.

The pressure is initialised consistently with the density and equation of state with constant reference, $\co$, and adiabatic index, $\Gamma$. The self-gravity potential is not initialised, and is computed by the code during the time-integration steps.

\subsection{Algorithm}

 We use the HLLC Riemann solver \citep{toro_restoration_1994} implemented in the code for the hydrodynamics part and a spectral solver for the Poisson equation implemented for this work and presented in App. \ref{app:PoissonSolver}. We also use the FARGO-like advection scheme \citep{masset_fargo_2000,stone_implementation_2010,mignone_conservative_2012} implemented in \idefix to speed up the integration.

\section{Results}

\subsection{Linear evolution}

In the isothermal case, we obtain a dispersion relation from Eqn.~(\ref{eq:linSW}) assuming $\sigma \propto e^{i\omega t}$ and in the case where $\xi_1$ is negligible

\begin{equation}
        -\omega^2 - 2q \Omega \frac{k_x k_y}{k^2} i \omega + \left[  k^2 \cs^2 - 2\pi G \Sigma_0 k + \kappa^2\left ( 1-q \frac{k_y^2}{k^2}\right)\right ] = 0\,.
        \label{eq:DispSW}
\end{equation}

We note that the axisymmetric case with $k_y=0$ yields the usual \cite{lin_spiral_1964} dispersion relation to find the Toomre instability criterion \citep{toomre_gravitational_1964}.

For a given spiral wave mode, there are in general, two complex roots to this dispersion relation

\begin{align}
    \omega_\pm &= -q\Omega \frac{k_x k_y}{k^2} i \nonumber\\
    &\pm \sqrt{-\left(q\Omega \frac{k_x k_y}{k^2}\right)^2 + \kappa^2\left( 1-q \frac{k_y^2}{k^2}\right) + k^2 \cs^2 -2 \pi G \Sigma_0 k}\,.
    \label{eq:roots}
\end{align}

\begin{figure*}
    \centering
    \includegraphics[width=\linewidth]{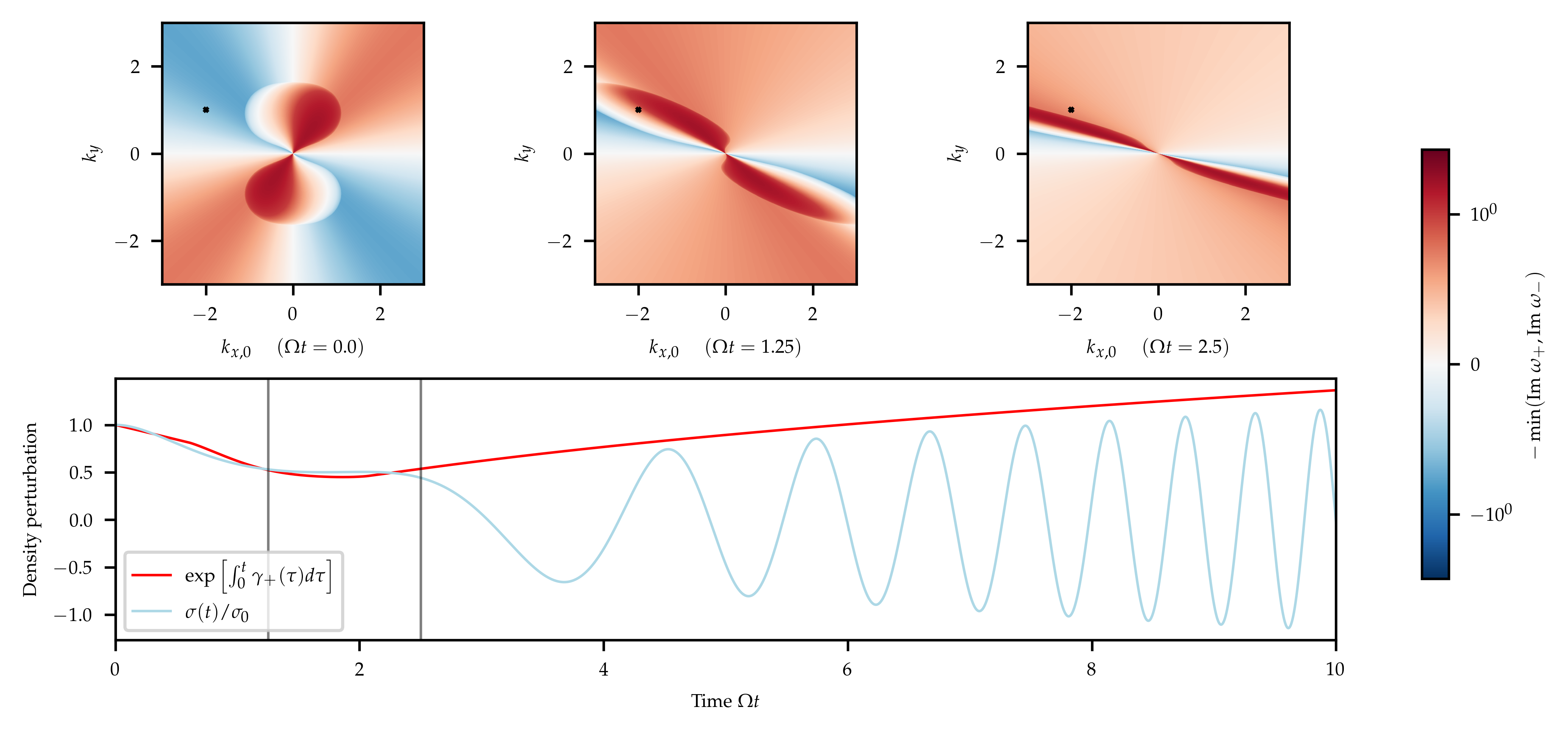}
    \caption{Time evolution of the growth rate of spiral density modes. \textbf{Top:} Maximum amplification root $\omega_\pm$ at different times. We note how the shear moves the most unstable region across the wave-vector plane. The times of these panels correspond $t=0$ and to the vertical grey lines of the bottom panel. \textbf{Bottom:} Example of the effective amplification factor with the $\gamma_{+}(t)$ factor for the mode $(k_{x,0}, k_y) = (-2,1)$, represented by a black "x" on the top panels. The light-blue curve is the semi-analytical solution for that mode.}
    \label{fig:ModeEvolution}
\end{figure*}

If one of them has a negative imaginary part, the corresponding $(k_{x0},k_y)$ spiral mode is unstable. The top panels of Fig.~\ref{fig:ModeEvolution} show the imaginary part of the roots of this dispersion relation at different times.

Because the $k_x$ wave vector is time-dependant, so are the roots and this will lead to change in behaviour over time such as transient amplification of certain modes. A better suited approximation is the Wentzel-Kramers-Brillouin-Jeffrey (WKBJ) approximation (see App. \ref{app:wkbj}). In our case, it can be shown that the solution will be of the form 

\begin{equation}
    \sigma^\text{WKBJ}_\pm(t) \sim A_\pm \sqrt{t} \exp \left[ i \frac{\cs q \Omega k_y}{2} t^2\right],
    \label{eq:wkbj}
\end{equation}
where $A_\pm$ is a constant. At large times, we have the asymptotic behaviour $\Im\ \omega_\pm \sim - \frac{1}{t}$, hence this is equivalent to having the a time-dependant growth rate related to the roots of the dispersion relation:
\begin{equation}
    \gamma_\pm(t) = - \frac{1}{2} \int_0^t \Im \omega_\pm (\tau) d\tau, 
    \label{eq:gamma}
\end{equation}
In the following, we refer to $\gamma_{\max} = \max ( \gamma_{-},\gamma_{+})$.

This effective growth rate takes into account past transient growth that would not be captured by merely evaluating the roots nor the WKBJ approximation. Indeed, the first approximation has no memory has it is analogous to some "instantaneous" growth rate, where as the second one only captures long term evolution and neglects any early time transient amplification or damping. The importance of capturing this initial transient growth has been shown in similar context (see for example \citealt{farrell_transient_1993,farrell_transient_2000,mamatsashvili_transient_2007}).

We have thus strike a golden mean with the approximation ${\vert \sigma(t) \vert \sim \sigma_0 \exp (\gamma_{\pm} t)}$ capturing both regimes.

On Fig. \ref{fig:ModeEvolution}, we see an example of a mode that is first damped, with initial $\Im \omega > 0$ and $\gamma_{\max} < 0$. This mode is then rapidly amplified when the strong amplification region encounters it. We note that at large time $\Im \ \omega_\pm \longrightarrow -q\Omega k_y$ meaning $\gamma_{\max}~\longrightarrow~\infty$. As the linear perturbation amplitude evolves $\propto \sqrt{t}$, linear theory is unlikely to describe the evolution of a spiral density mode very long, let alone up to infinity, but this is a good motivation to carry out this work to a non-linear regime as we do in the next section.

\begin{figure}
    \centering
    \includegraphics[width=\columnwidth]{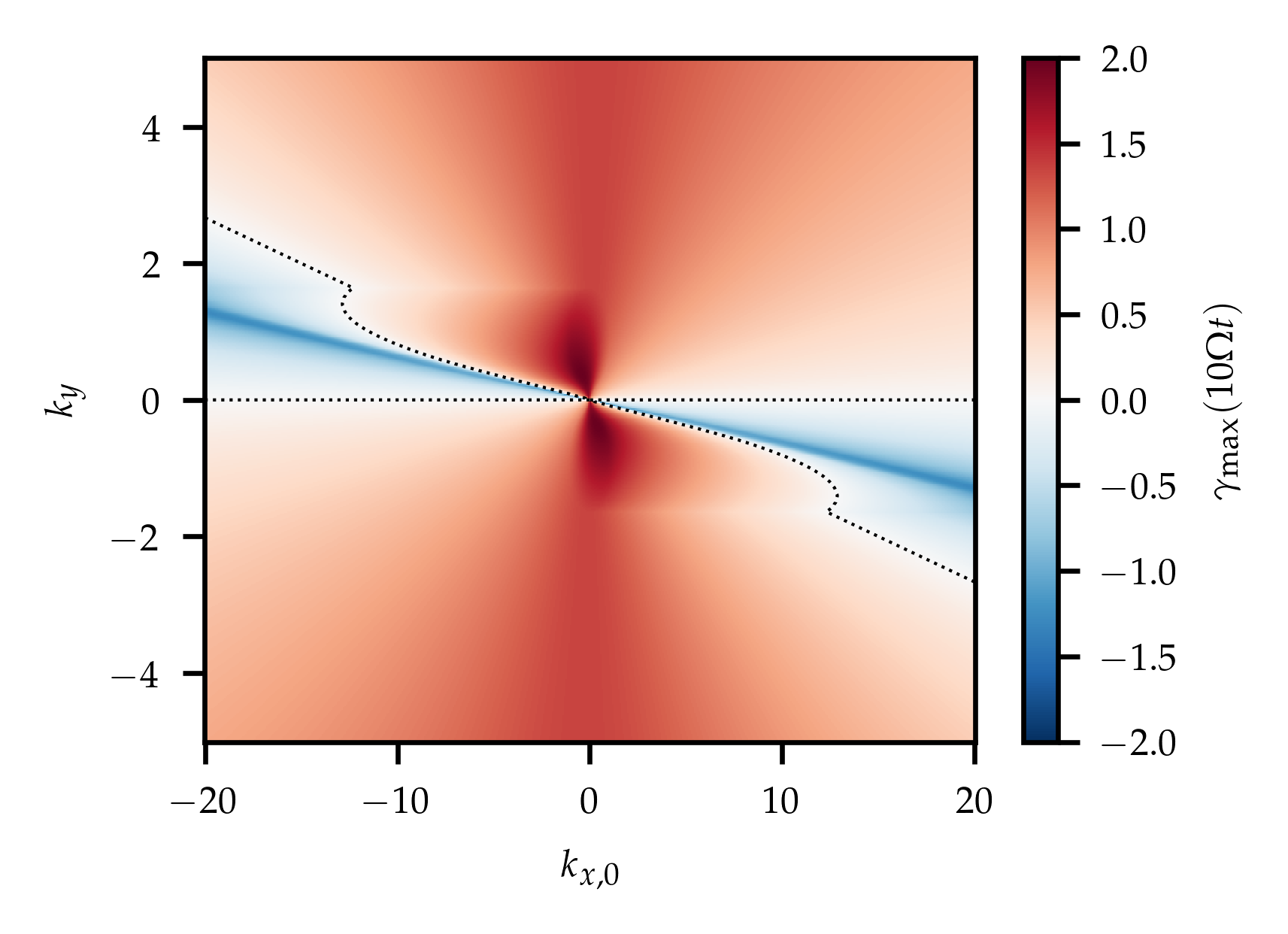}
    \caption{Contour of the imaginary part of the roots of the effective growth rate after 10 local orbital times for $Q = 3/2$. Positive values (red) correspond to amplified modes, negative values (blue) correspond to damped modes, the black dotted line corresponds to constant amplitude modes. We note that the amplification region is smeared out along the $k_x$ direction like shown on Fig. \ref{fig:ModeEvolution}. We note that if one were to evaluate $\gamma_{\max}$ at later time, only a region of decreasing size would be damped.}
    \label{fig:DispRel}
\end{figure}

From Eqn.~(\ref{eq:roots}) and Fig. \ref{fig:DispRel}, the most unstable modes are large-scale modes, with a small wave number -- similar to the \cite{lin_spiral_1964} axisymmetric dispersion relation, which has the most unstable mode for $k_x = \kappa/(\cs Q) \approx H/Q$. Here, we also have a $\propto \Sigma_0$ dependency on the most unstable wave number. We note that for $Q \longrightarrow\infty$, the most unstable mode is for $(k_x,k_y) \longrightarrow (0,0)$.
We note from Eqn.~(\ref{eq:roots}) that the dispersion relation does not behave smoothly at $(0,0)$, and it should be noted that the uniform mode $(k_x,k_y) = (0,0)$ is stable. Physically, this mode is the average density and is always conserved.

More interestingly, we note that the growth rate of the most unstable mode does not vanish when $\Sigma_0 \longrightarrow 0$. Instead, we compute $\min_{\bm{k} \in \mathbb{R}^2} (\Im \ \omega_\pm)\longrightarrow -\sqrt{2}$ for a Keplerian disc. This means that this instability is never killed at low density. As a matter of fact, in the previous derivation it might seem as if self-gravity does not play a role in the evolution of a spiral wave density mode. It is true that the long term amplification of the WKBJ solution is not different for a non-self-gravitating disc, however the self-gravity will strongly increase the transient amplification of each mode (compare for example Figs. \ref{fig:SWnoSG} and \ref{fig:SWSG}), hence the importance of integrating the growth rate over time as we do in Eqn. (\ref{eq:gamma}).

We compare the results from the linear stability analysis to full non-linear simulations in the following.

\subsection{Numerical Results}

\begin{table}[!ht]
	\centering
	\caption{List of simulations presented in this paper}
	\label{tab:simulations}
	\begin{tabular}{cc|cc|cc}
		\hline
		$\Sigma_0 = Q^{-1}$ & $\Gamma$ & $\Sigma_0 = Q^{-1}$ & $\Gamma$ & $\Sigma_0 = Q^{-1}$ & $\Gamma$ \\
		\hline
        2/3 & iso & 2/3 & 1.1 & 2/3 & 5/3  \\
		0.80 & iso & 0.80 & 1.1 & 0.80 & 5/3   \\
		0.90 & iso & 0.90 & 1.1 & 0.90 & 5/3  \\
		0.95 & iso & 0.95 & 1.1 & 0.95 & 5/3 \\
        0.99 & iso & 0.99 & 1.1 & 0.99 & 5/3 \\
		\hline
	\end{tabular}
\end{table}

Table~\ref{tab:simulations} summarizes the 15 simulations performed for different values of $\Sigma_0 = Q^{-1}$ and adiabatic index $\Gamma$. In this table, iso refers to isothermal simulations for which the energy equation is not solved, and the equation of state reduces to the $\Gamma = 1$ case. All simulations share the same initial condition, which is a spatially uniform profile with constant Keplerian shear rate $q = 3/2$.

In the following, we refer to the simulations with the following naming convention: \texttt{GXXXX\_SYYYY} where \texttt{XXXX} is the rounded adiabatic index $\Gamma$ and \texttt{YYYY} is the rounded background density $\Sigma_0$. The isothermal simulations have the names \texttt{ISO\_SYYYY} with same the convention for the density.

\subsection{Spiral waves}

\subsubsection{Growth regimes}

\begin{figure}[!ht]
    \centering
    \includegraphics[width=0.92\columnwidth]{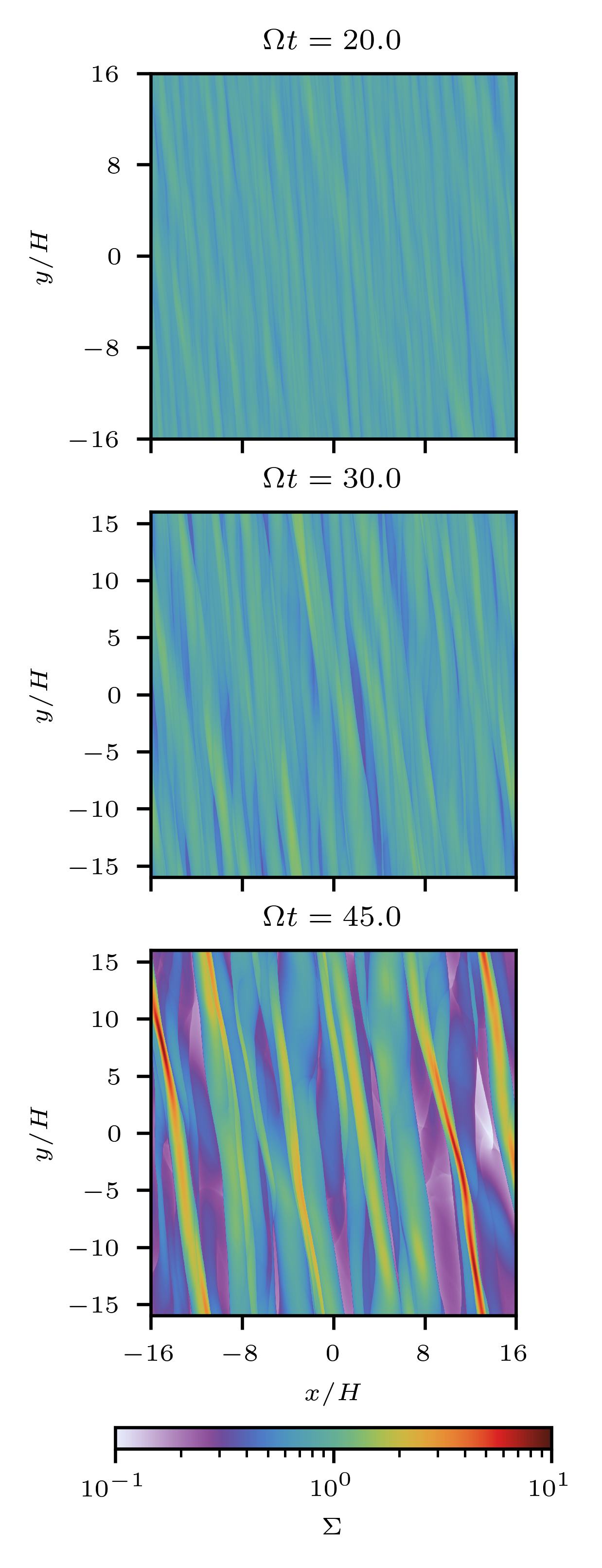}
    \caption{Density map of snapshots of the \texttt{G1.10\_S0.80} runs at different times. Before $\Omega t<10$ spiral arms have too low contrast to be seen with this normalisation. After $\Omega t>45$, the contrast is also increased when the spiral arms collapse. On the first panel, the spirals are in the linear growth regime; for the last two panels, where the spiral arms overlap in some places, they are in the non-linear (interaction) regime.}
    \label{fig:snaps}
\end{figure}

In all simulations, the spiral wave evolution can be split in two regimes. First comes the linear growth regime, matching the linear theory. During this phase, non-axisymmetric modes grow in a similar fashion to that of an amplified oscillator. Then, when the amplitudes of the non-axisymmetric modes become non-negligible with respect to the axisymmetric mode, we reach a non-linear amplification regime. Physically, it corresponds to when the spiral arms have a strong density contrast with respect to the mean flow and start interacting with each other.

To study individual spiral modes, we look at the shearing-wave decomposition of the density field. This can be done with a shifted Fourier transform, analogous to the SAFI (Shear Advection by Fourier Interpolation -- \citealt{johansen_zonal_2009}) method presented in App. \ref{app:PoissonSolver}. This amounts to projecting the density on the functional basis of equation (\ref{eq:SWbasis}).

\begin{figure}
    \centering
    \includegraphics[width=\columnwidth]{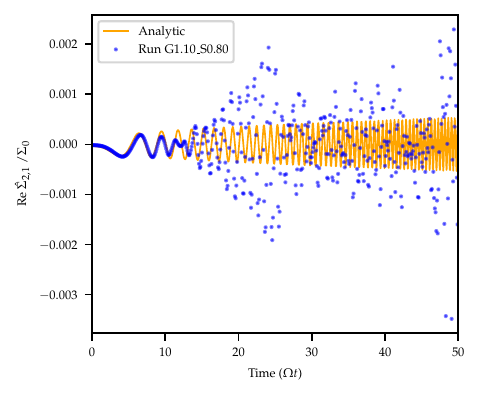}
    \caption{Time evolution of the shearing-wave modes $(\ell,m)=(2,1)$ of \texttt{G1.10\_S0.80}. The blue dots measure the shearing-wave mode amplitude, and the orange solid line corresponds to the analytical solution of Eqn.~(\ref{eq:linSW}) for the corresponding mode.}
    \label{fig:GrowthPhases}
\end{figure}

\begin{figure*}
    \centering
    \includegraphics[width=\textwidth]{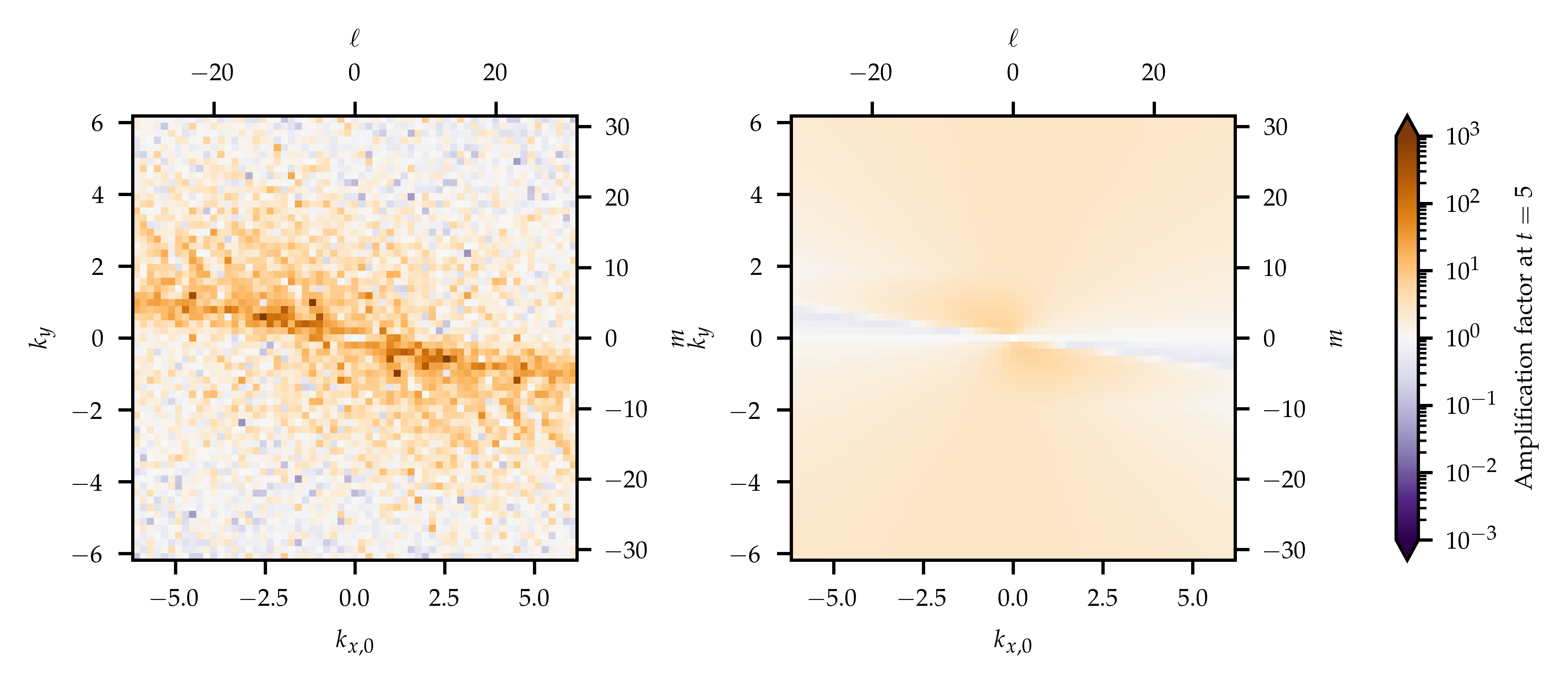}
    \includegraphics[width=\textwidth]{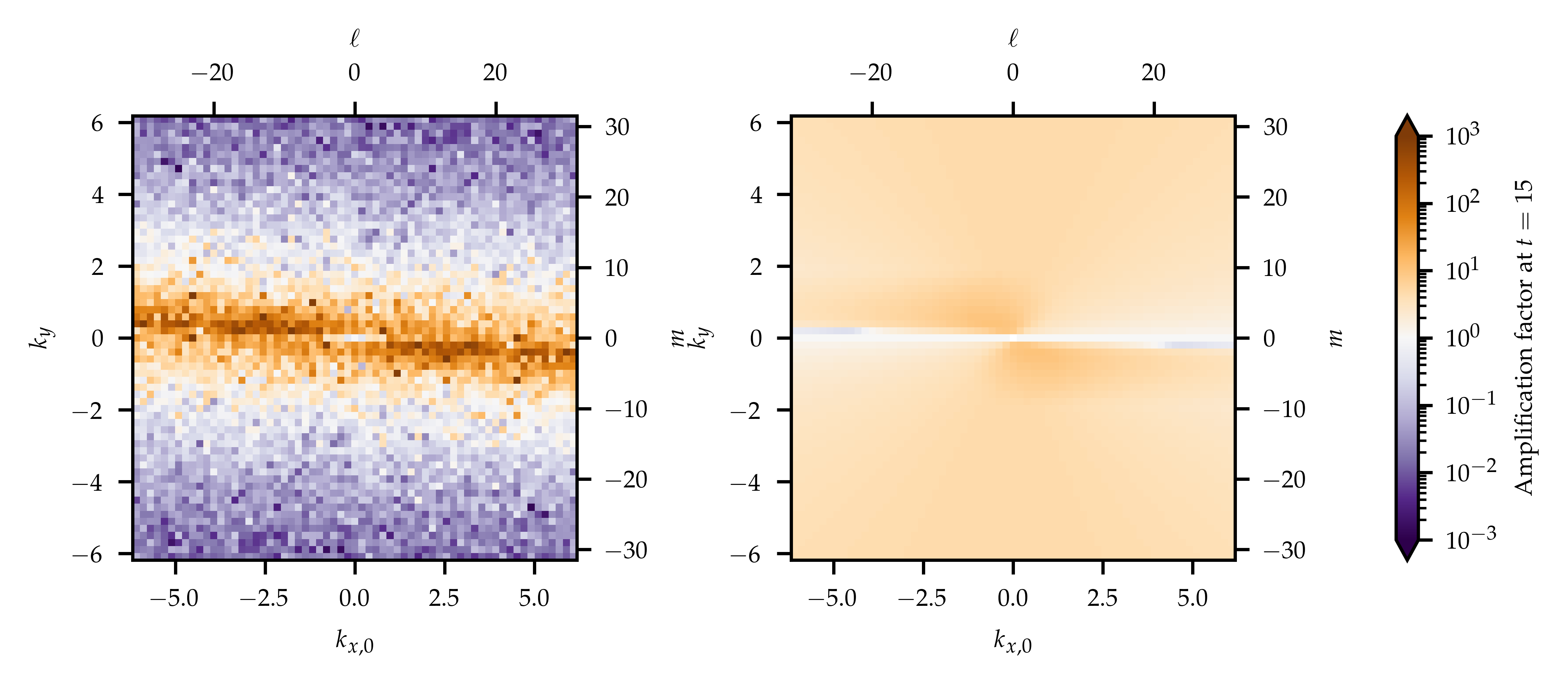}

    \caption{Mode amplification for the \texttt{G1.10\_S0.80} simulation. The axis scales are given both in terms of wave vector $k_{x,0},k_y$ and corresponding wave number $\ell,m$ for the box size of the simulation. \textbf{Left:}  Measured amplification factor from the simulations. \textbf{Right:} Analytic amplification factor $\sim \sqrt{t}$ at large times computed from Eqn. \ref{eq:gamma}. We show the numerical quantity for the intermediate $\Gamma=1.1$ simulation, but this analytic amplification factor is computed under the isothermal assumption, as $\Gamma\neq1$ can not be linearised. \textbf{Top:} At $\Omega t = 5$. \textbf{Bottom:} At $\Omega t =15$.}
    \label{fig:CompToAnal}
\end{figure*}

Figure \ref{fig:GrowthPhases} shows the time evolution of the mode ${(\ell,m)=(2,1)}$ of the simulation \texttt{G1.10\_S0.80}. For the first 10 orbital timescales, we observe that the evolution of the spiral mode matches almost exactly the linear solution. We note that the small difference comes from the $\Gamma\neq 1$ non-linearity. Appendix~\ref{app:BC} shows perfect agreement for the isothermal case. At later times, however, its evolution departs from it. At around $\Omega t = 15$, we see a very strong growth due to non-linear interactions between the spiral modes. For this simulation, a similar change of regime is observed across all modes. For other simulations, a similar behaviour is observed, but the time at which it happens varies from simulation to simulation. The latest phase, $\Omega t > 30$, corresponds to saturation of the spiral modes.

In order to quantify the departure from linear theory, we compute an effective amplification factor $\vert \sigma_{\ell,m}(t)/\sigma_{\ell,m}(t=0)\vert$ from the shearing wave decomposition. From Eqs. (\ref{eq:wkbj}-\ref{eq:gamma}) the linear theory prediction for this quantity has asymptotic behaviour $\sim e^{\gamma_\pm(t)} = \sqrt{t}$.

Figure \ref{fig:CompToAnal} shows the measured and linear theory amplification for simulation  \texttt{G1.10\_S0.80} for the modes $\ell,m \in \llbracket -31,31 \rrbracket$. This quantifies the departure from linear theory at large times both at large and small scales. We measure for all simulations that the modes which undergo strongest amplification have amplification factors reaching of $10^4$ for all simulations after $\Omega t = 15$, this exceeds the linear amplification factor by more that $\times 1000$. We discuss mode selection, that is the dependency of the amplification factor with wave number/vector, in the following section. 

All simulations have qualitatively a similar behaviour but the amplification does vary from one to the other. Simulation with higher density have overall stronger amplification for all modes, with the small scale modes being less damped. Changing the thermodynamics also affects the amplification: simulation with lower $\Gamma$ also show stronger amplification across all modes.

In all simulations, we observe these three phases:
1. Linear growth for around ten orbits. 2. Rapid non-linear growth. 3. Saturation. We note that the saturation phase can not be precisely captured in the simulations presented here. In the non-isothermal simulations, this is because there is continuous shock heating from the spiral wave that heats up the disc. This heating is not being compensated for by any cooling function and will gradually destroy the spirals by increasing the disc temperature. In the isothermal case, because there is no shock heating happening at short time-scales, over-densities undergo runaway collapse.

\subsubsection{Mode selection}

As we initialise the runs with white noise, all $m>0$ modes start with similar amplitude. However, not all modes grow at the same rate. Consistently with linear theory, the largest scale modes grow the fastest and end up dominating the density profile.

\paragraph{Shearing wave modes} On Fig. \ref{fig:CompToAnal}, we can clearly see a strong mode selection after $\Omega t \geq 15$. The modes with $\vert m \vert \lesssim 15$ are amplified, with amplification factor $>1$ (brown on the figure), whereas the smallest scales modes are damped, with amplification factor $< 1$ (purple on the figure). We note that the amplified modes coincide with those with the strongest linear theory amplification. Yet, linear theory predicts that all mode mode should be amplified and here we observe damping of small scales modes. We note that the behaviour of small scales modes should be regarded with caution as they are less resolved. Modes with $\ell, m > 32$ are resolved by less than 64 point per wavelength ; comparing to App. \ref{app:PoissonSolver}, we see that for such resolution, numerical damping can be expected.

\paragraph{Azimuthal modes}

The shearing-wave mode decomposition presented above is very useful to measure single modes and compare their evolution to linear theory. We have now established that there is strong non-linear growth and we want to link to the more global dynamics, the $x$-direction is thus of lesser interest here. We thus look at the azimuthal direction only Fourier decomposition. We keep the name $m$ for the azimuthal wave number, as they fundamentally capture the same thing. To obtain a one-dimensional quantity, we average over the $x$ direction. We write with a tilde the quantity obtained in that way.

\begin{figure}
    \centering
    \includegraphics[width=\columnwidth]{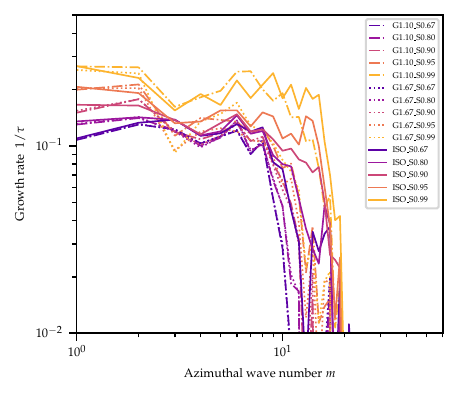}
    \caption{Growth rate of the azimuthal spiral modes, assuming $\vert \tilde{\Sigma}_m \vert \propto e^{t/\tau}$, as a function of their wave number for $\Omega t < 15$. We voluntarily clip the plot at $\tau = 100$, as a larger growth timescale ($\gg 15$) cannot be captured properly.}
    \label{fig:GRm}
\end{figure}

Initially, the $\tilde{\Sigma}_m$'s grow exponentially in amplitude. We measure a growth rate by fitting its amplitude with an exponential growth $\propto e^{t/\tau}$. Figure \ref{fig:GRm} shows how the largest scale modes grow much faster than the small scale modes. On this figure we see that modes up to $m \approx 10 -20$ grow very fast, with $1/\tau \approx 10^{-1}$ whereas the small scales modes have much smaller $1/\tau \approx 10^{-8}$. More precisely, the plot shows us that in all simulations it is the $m=1$ or $m=2$ mode that grows the fastest.

Figure \ref{fig:GRm} also shows that only the modes with wavelength $\lambda_m = L_y/m \gg H$ show growth. With our grid and units, $\lambda_m = H$ for $m=32$. On the figure, we see that only the modes with $m < 20$ show significant growth, with slight variations with initial density and adiabatic index. This is consistent with the observations of \cite{gammie_nonlinear_2001}.

\begin{figure}
    \centering
    \includegraphics[width=\columnwidth]{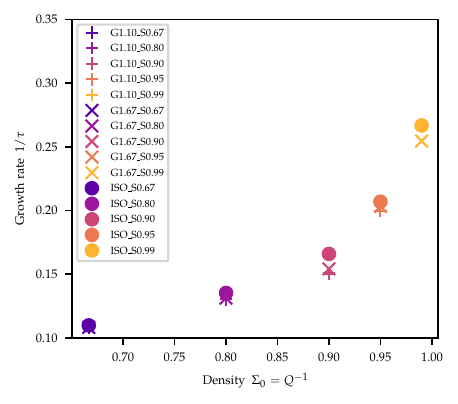}
    \caption{Growth rate of the first azimuthal spiral mode ($m=1$) for all simulations as a function of the initial density for $\Omega t < 15$.}
    \label{fig:GR1}
\end{figure}

Figure \ref{fig:GR1} shows how the growth rate of the $m=1$ mode changes with the parameters of the simulations presented in this paper. We see only little variation with thermodynamics, with the isothermal runs growing the fastest. This can be understood as they do not heat and have lower pressure to contract self-gravity. The dependency on the initial density, however, is much stronger. We see that the growth rate increases over-linearly with the initial density. We note that even in the lowest density disc probed, we measure very rapid growth with $\tau \lesssim 10$. The fact that we observe only little variation with thermodynamic is due to the fact that we are only probing the initial growth of the spirals. Runs with different thermodynamics saturate in different ways, which can not be properly captured in this work, as further discussed in Sect. \ref{sec:Discussion}.

\begin{figure}
    \centering
    \includegraphics[width=\columnwidth]{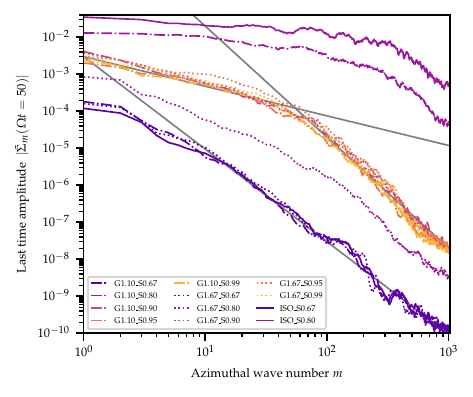}
    \caption{Density power spectrum with respect to the azimuthal wave number at the end of the simulations. The grey lines are the power laws described in the text.}
    \label{fig:AmpLate}
\end{figure}

At later times, the exponential growth poorly describes the evolution of the spiral modes. In all simulations, at the end of the integration time, at $\Omega t = 50$, the mode $m=1$ still dominates. Figure \ref{fig:AmpLate} shows how the amplitude $\tilde{\Sigma}_m$ decreases with azimuthal wave number at this time. We note that the high-density isothermal simulations are not shown on this plot as they undergo runaway collapse before this time and crash due to the absence of shock heating.

We measure a steep decay in the density power spectrum of $\propto m^{-5/2}$ for the lowest density simulations. Most other simulations that reach a highly non-linear state ($\Sigma_0 \geq 0.9$) have a broken power-law spectrum with $\propto m^{-4/5}$ at low wave-number and  $\propto m^{-3}$ at small scales. The grey lines of the plot represent these power-law behaviours and were not fitted but adjusted only by eye. The simulations with $\Sigma_0 = 0.8$ appear to be outliers. For the isothermal one (solid line), this is because it is on the verge of undergoing runaway collapse, and all modes are very rapidly growing at that time. The $\Gamma = 1.1$ is different because it is reaching the peak of it non-linear state at this time (see Fig. \ref{fig:AlphaAll} for example) while the other simulations with $\Gamma = 1.1$ with higher density had already reached it before and are decaying because of the absence of cooling.

\subsection{Spiral-driven accretion}

In addition to producing local over-densities, these spiral waves also contribute to the angular momentum transport in the accretion disc. From local shearing sheet simulations, we can not straightforwardly compute an accretion rate, but it is possible to measure the local stresses.

Here, their the total stress can be decomposed into two contributions, the Reynolds stress and the gravitational stress.

\begin{equation}
     W_{x,y} = \Sigma v_x' v_y' + \frac{g_x g_y}{4 \pi G}
\end{equation}
where $\bm{v}'$ is perturbation velocity around the sheared back ground $\bm{v}'= \bm{v} +q\Omega \bm{e}_y$, with $\bm{v}$ the total velocity and $\bm{g} = -\nabla \phi$ is the gravitational field. 

These stresses can be converted into an $\alpha$ angular transport parameter \citep{shakura_black_1973}, by averaging and dividing by the pressure.

\begin{align}
        \alpha_\mathrm{tot} &= \alpha_\mathrm{R} + \alpha_\mathrm{G} \\
        &= \frac{\langle \Sigma v_x' v_y' \rangle}{\langle P \rangle} + \frac{\langle g_x g_y \rangle }{4 \pi G \langle P \rangle}
\end{align}

where $\langle \dots\rangle$ is the average over the entire simulation domain.

\begin{figure*}
    \centering
    \includegraphics[width=\linewidth]{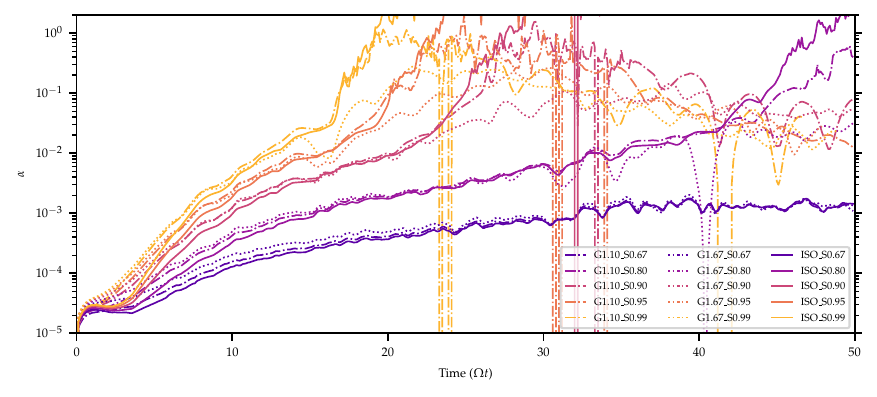}
    \caption{Time evolution of the angular transport parameter for all simulations. We note that at the strongest of the collapse, some simulation show negative $\alpha$ values, these are only transient features.}
    \label{fig:AlphaAll}
\end{figure*}

Figure \ref{fig:AlphaAll} shows the time evolution of $\alpha$ across all simulations. On this figure, we observe the transient accretion regime that ensues from the spiral emergence.

\begin{figure}
    \centering
    \includegraphics[width=\columnwidth]{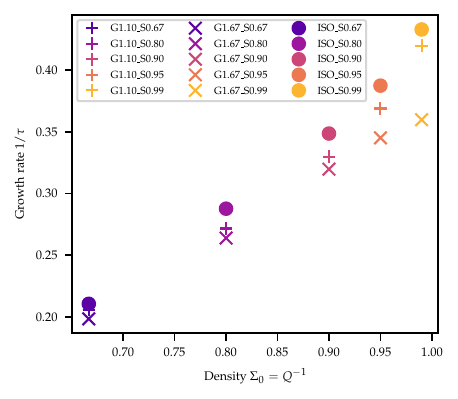}
    \caption{Growth rate of the angular transport parameter $\alpha$ for all simulations for $\Omega t < 13$.}
    \label{fig:AlphaGR}
\end{figure}

We observe that the initial growth rate (for $\Omega t < 13$) does not depend much on the thermodynamics of the disc. Indeed, discs with different $\Gamma$ but similar initial density behave very similarly. However, the initial density changes the growth rate dramatically. The initial growth phase is exponential, and we can extract a growth rate $1/\tau$ by fitting the evolution for the first 13 orbital timescales. This growth rate is plotted on figure \ref{fig:AlphaGR}. 

As we can better see in this figure, there is still a small dependency on the disc thermodynamics. Indeed, discs that heat up less rapidly (low $\Gamma$ and isothermal) show more rapidly growing $\alpha$ as pressure counteracts gravity less efficiently.

We note that this initial growth phase is only a transient regime and the $\alpha$ values measured there can not be extrapolated to a possible long-standing spiral accretion regime. However, this rapid growth shows the disc response to spiral wave perturbations. This means that a global disc being fed matter from an external source, for example large-scale accretion streamer, or the infalling initial cloud, as seen in \cite{mauxion_modeling_2024} will have a transient accretion response. Here we see that both the growth rate and the 'final' $\alpha$ value depend on the initial density, meaning that an influx of mass will trigger an increased accretion phase with amplitude depending on the amount of matter. A varying $\alpha$ value can be linked to outbursting systems like FU Ori (\citealt{hartmann_FU_1996,bourdarot_FUOri_2023}, see also for example, the Disc Instability Model, and \cite{hameury_review_2020} for a review)

In turn, this effective viscosity, captured by the $\alpha$ parameter, contributes to viscous heating of the disc and will change its luminosity, much like in compact binary systems.

\begin{figure}
    \centering
    \includegraphics[width=\columnwidth]{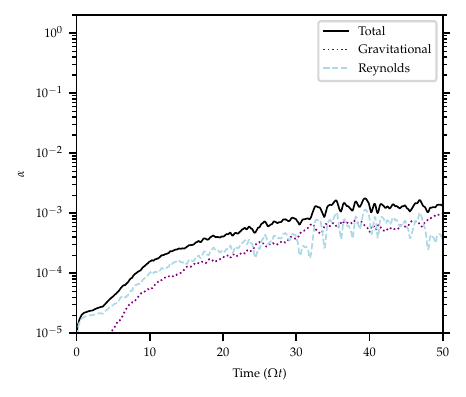}
    \caption{Angular momentum transport parameter of the simulation \texttt{G1.10\_S0.67}. The solid black line is $\alpha_\mathrm{tot}$, the purple dotted line is $\alpha_\mathrm{G}$ and the dashed blue line is $\alpha_\mathrm{R}$.}
    \label{fig:Alpha0.67}
\end{figure}

\begin{figure}
    \centering
    \includegraphics[width=\columnwidth]{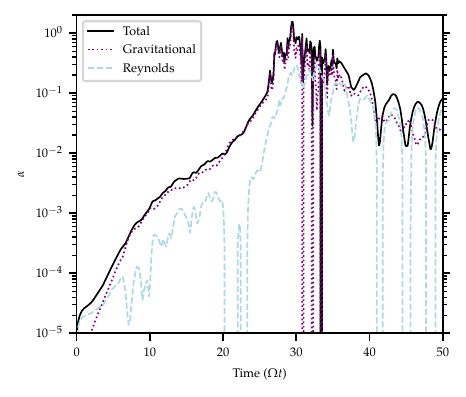}
    \caption{Angular momentum transport parameter of the simulation \texttt{G1.10\_S0.90}. The solid black line is $\alpha_\mathrm{tot}$, the purple dotted line is $\alpha_\mathrm{G}$ and the dashed blue line is $\alpha_\mathrm{R}$.}
    \label{fig:Alpha0.90}
\end{figure}

As mentioned above, it is possible to measure the stresses that make up $\alpha$ individually. Figures \ref{fig:Alpha0.67} and \ref{fig:Alpha0.90} show this decomposition for the simulations \texttt{G1.10\_S0.67} and \texttt{G1.10\_S0.90}. On the first one, we see that initially $\alpha$ is dominated by the hydrodynamical stresses (Reynolds) and after $\Omega t > 30$, both the gravitational stress and the hydrodynamical stress become comparable. We obtain some sort of weak gravito-turbulent regime, which is dominated by self-gravitating spiral shocks. This simulation never develops strong gravito-turbulence like \cite{gammie_nonlinear_2001} as here the initial Toomre criterion of the disc is $Q = 1.5 > 1$. We note however, that we obtain non-negligible $\alpha \approx 10^{-3}$ even in a Toomre stable disc by spiral-driven accretion only. We observe this for all simulations with this low initial density. These simulations with low initial density are the only ones to reach an almost steady state. This is well illustrated by their plateauing $\alpha$ on Fig. \ref{fig:AlphaAll}.

Figure \ref{fig:Alpha0.90} shows the $\alpha$ parameter decomposition for a simulation with a higher initial density. There, the gravitational stresses dominate by a large amount after only a few orbital timescales. We observe similar behaviour for all simulations with $\Sigma_0 \geq 0.80$. The $\alpha$ then rapidly rises to reach values of 1, proof that a highly non-linear regime is reached. The decay observed after $\Omega t > 30$ is a consequence of the unbalanced heating of the disc. This will be further discussed in the following section.

\section{Discussion}
\label{sec:Discussion}

\subsection{Global vs Local Picture}
The simulations and analytics presented here all make the assumption of locality for the spiral wave in the form of the Hill approximation. We observe that the largest scale spiral modes dominate the density. This suggests that in actual global discs, the dominant modes will have only a few spiral arms as has been observed \citep{fukagawa_spiral_2004,muto_discovery_2012,wagner_discovery_2015,benisty_asymmetric_2015,huang_disk_2018}. However, because the formalism developed here lacks some global properties, such as curvature terms, and inside-outside dissymmetry, these results can not be quantitatively translated to global discs. The fact that the largest mode dominated hints towards the fact that a study of self-gravitating spiral waves reaches the limits of the local approximation. 

\subsection{Life-span}

As mentioned above, in this work, we are not able to capture any long-standing saturated spiral wave regime. The reason for this is that we do not include any form of cooling to counteract spiral shock heating. In the isothermal simulations, even though the temperature is kept constant, we still can not study this regime. This is because the shock heating due to spirals is not captured and takes place on very short timescales. As a consequence, a local over-density, provided it is dense enough, will undergo runaway collapse down to scales lower than the resolution of the simulation, and will not be counteracted by increasing pressure from shock-heating.

A better treatment, including a cooling function, would allow us to study the balance between short-timescale heating due to spiral shock and the cooling. Knowing the life span of the spirals is crucial as it would tell us how likely we are to observe them in actual systems. In this work, we have shown that spiral wave form on a very short time-scale, of around 10 dynamical time-scales. If they are long-lived, it means that there is only an extremely short time window when they do not exist.

We have also seen that these spiral shocks drive significant accretion through the measurement of the Reynolds and gravitational stresses. In a global disc, a high level of accretion will change its dynamics, and can not be sustained indefinitely, as the disc is eventually depleted.

\subsection{Spirals as dust traps}

The life span of the spiral structures is also crucial when one takes into account the presence of dust in the accretion disc. They create local over-density regions that could host a large amount of dust accumulation, for example, through the streaming instability (\cite{youdin_streaming_2005} and following papers) or effects due to the vertical structure \citep{lehmann_impact_2022}. If the spirals are long-lived, they may host planetesimal formation. However, this would require a large amount of dust to accumulate, at which point the gravitational influence of dust can no longer be neglected. This accumulation of dust may destroy the spiral structures and, in turn, prevent a runaway dust accumulation (see \cite{lesur_PPVII} for a review).

\section{Conclusion}

In this work, we have shown that self-gravitating discs are unstable to non-axisymmetric mode growth in regimes that are usually considered stable through the axisymmetric Toomre stability analysis. More precisely, we have shown 1) that transient growth of spiral density waves needs to be taken into account when linear theory, 2) that this transient growth is rapidly superseded by very strong non-linear growth with growth rates $> 0.1 \Omega$.

These spiral waves are first well described by their linear evolution. The evolution of their amplitude is similar to that of an amplified oscillator. And similarly to the axisymmetric regime, whose most unstable mode is for $k_x = H/Q$, here, it is the largest-scale spiral mode that grows the fastest, with most unstable non-axisymmetric mode also $k_y \propto 1/Q$. We confirmed this theoretical prediction with high-resolution non-linear simulations.

After a short time (10 local orbital time scales) their amplitude has grown enough that the linear approximation poorly describes their behaviour. In this regime, non-linear interaction leads to the even faster growth of the large-scale modes.

We showed that the most dramatic control parameter for the growth of these spirals is the initial density (or equivalently, the corresponding axisymmetric Toomre parameter). The higher the density, the faster the growth of spirals. The thermodynamics (adiabatic index) of the gas plays a somewhat secondary role in the spiral wave growth. The faster the disc heats up (with high $\Gamma$), the slower the spiral over-density grows as they are being counteracted by non-linearly increasing pressure.

These spiral waves also come with increased torque on the disc. High-density discs are dominated by gravitational stresses that may lead to increased transient accretion regimes. Low-density discs still have equal contributions of hydrodynamic and gravitational stresses and can sustain regimes of $\alpha \approx 10^{-3}$.

This work only focused on the initial growth of the spiral waves, and a more realistic treatment of thermodynamics needs to be introduced if one hopes to study the long-term evolution of the spiral waves and possible further collapse.

\begin{acknowledgements}
  This work was funded\,\footnote{Views and opinions expressed are however those of the author(s) only and do not necessarily reflect those of the European Union or the European Research Council. Neither the European Union nor the granting authority can be held responsible for them.} by the European Union (ERC-CoG, \textsc{Epoch-of-Taurus}, No. 101043302). The simulations and data reduction were performed on the TU Dresden/ZIH cluster. The authors gratefully acknowledge the computing time made available to them on the high-performance computer at the NHR Center of TU Dresden. This center is jointly supported by the Federal Ministry of Research, Technology and Space of Germany and the state governments participating in the NHR (\url{www.nhr-verein.de/unsere-partner}). Analysis and reduction of simulation data were performed using SciPy \citep{2020SciPy-NMeth}, NumPy \cite{harris2020array}, and Matplotlib \citep{Hunter:2007}. \idefix and the spectral solver implemented for this work make extensive use of the \kokkos portability tools \citep{9502936,9485033,CarterEdwards20143202}.
\end{acknowledgements}

\bibliography{bib}
\bibliographystyle{aa}

\appendix
\section{WKBJ solution}
\label{app:wkbj}

We note that computing a growth rate from the linear shearing wave differential equation or the associated "instantaneous" dispersion relation is not straightforward as it is not with a constant coefficient (we would have a time-dependent growth rate $\omega(t)$). Similarly, finding an analytical solution is possible by rewriting the equation as a vector linear first-order ordinary differential equation (ODE) $\dot{X} = \mathbf{M} X$, with $X = (\dot{\sigma},\sigma)$. The analytical solution is 

\begin{equation}
    X(t) = \operatorname{OE}[\mathbf{M}](t) X_0 = \mathcal{T}   \left \{ e^{\int_0^t \mathbf{M}(s) ds }\right \} X_0 ,
\end{equation}

\noindent where $X_0$ is the initial condition and where $\operatorname{OE}$ or $\mathcal{T}   \left \{ e^{\dots} \right \}$ is the time-ordered exponential. This operator is required as the matrix $\mathbf{M}$ is a function of time and does not commute with itself at different times. We note that the existence of this exact solution is unfortunately of little help, as it does not provide us with simple parameters to control the solution, such as a constant growth rate, as is the case with a constant-coefficient first-order ODE. Indeed, computing this time-ordered exponential is not much simpler than integrating the ODE with a readily available ODE solver.

Instead we can compute an approximate solution with the WKBJ approximation as follows. For a general homogeneous linear ODE of order 2 of the form 
\begin{equation}
	\ddot{y} + a(t) \dot{y} + b(t) y = 0,
\end{equation}
the general WKBJ solution is 

\begin{align}
        y_\pm(t) &=  A_\pm \left[b(t) - \frac{a^2(t)}{4} - \frac{\dot{a}(t)}{2}\right]^{-1/4} \nonumber\\
        & \times\exp\left(\int_{0}^t \left[-\frac{a(s)}{2} \pm \sqrt{-b(s) + \frac{a^2(s)}{4} + \frac{\dot{a}(s)}{2}}\right] ds\right),
\end{align}
where $A_\pm$ are constants to be determined from the initial condition.
In our case, we have the following asymptotic behaviours at large time:

\begin{align}
    a(t) &\sim \frac{2}{t},\\
    \dot{a}(t) &\sim -\frac{2}{t^2},\\
    b(t) &\sim (\cs q \Omega k_y t)^2.
\end{align}

From these, we obtain the asymptotic behaviour of the WKBJ solution for the amplitude:

\begin{equation}
    y_\pm(t) \sim \tilde{A}_\pm \sqrt{t}\ \exp \left(i\frac{\cs q \Omega k_y}{2} t^2\right)
\end{equation}

Note that the actual solution is a linear combination of both $+$ and $-$ solution.

\section{Spectral Poisson solver}
\label{app:PoissonSolver}
In this work, we use the \idefix code to produce 2D simulations of a razor thin disc. The solver implemented in this code, uses a spectral method implemented for this work. To ensure performance portability, the \idefix implementation of this solver uses the \texttt{Kokkos-fft} library \citep{asahi_kokkosfft_2025}.

\subsection{SAFI-like method}
To compute the self-gravity potential, we modify the SAFI method introduced by \cite{johansen_zonal_2009}. The main difference with their method, is that they only use the Fourier-space for the interpolation of the shear advection. Here, we use a fully spectral method. The steps corresponding to the SAFI method are steps (\ref{eq:SolverS1}) and (\ref{eq:SolverS3}) in the following.

The reason for using such an interpolation methods is that the shearing sheet (or box) geometry does not allow a straightforward solving of Poisson equation, as it is shear-periodic not just periodic. In order to take this into account, we first do a Fourier transform along the $y$ direction. 

\begin{equation}
    \hat{\rho}^{(y)}_\text{unsheared} = \hat{\rho}^{(y)} e^{is}\,,
    \label{eq:SolverS1}
\end{equation}

where $s = S t_\text{p} x k_y$, and $t_\text{p}$ is the periodic time, that is the usual simulation time $t$ modulo the shear-period $T_\text{shear} = 2/S $.

We then do a Fourier transform along $x$ and we solve the Poisson equation for this unsheared density, with some kernel function $K$ (the Fourier transform of the Green's function).

\begin{equation}
    \hat{\phi}_\text{unsheared} = K \ \hat{\rho}_\text{unsheared} \,.
    \label{eq:SolverS2}
\end{equation}

To go back in the real space, we preform an inverse Fourier transform along $x$, and we then reshear the potential.

\begin{equation}
    \hat{\phi}^{(y)} = \hat{\phi}_\text{unsheared}^{(y)} e^{-is}\,.
    \label{eq:SolverS3}
\end{equation}

Finally, we perform an inverse Fourier transformation along $y$.
In total, this procedure includes four 1D Fourier transforms; the shear translation does not add more transforms compared to a regular periodic geometry.

\subsection{Kernel used}

In the case of a razor thin disc $\rho \propto \Sigma \delta(z)$, the appropriate kernel to solve Poisson equation is 

\begin{equation}
    K_\texttt{dirac}(k) = - \frac{2\pi G}{\vert k \vert}\,.
\end{equation}

The code comes with two default kernels relevant for usual disc geometries. We provide the \texttt{normal} kernel

\begin{equation}
    K_\texttt{normal}(k) = -4 \pi G k^2\,,
\end{equation}

as well as the \texttt{dirac} kernel above. The code also supports user-defined kernels that can be defined in the setup files without tinkering with the code base. We refer the reader to the online documentation.

\subsection{Performance benchmark}

We compare the performance of the solver implemented here with the iterative solver implemented in \idefix \citep[cf.][for details]{mauxion_modeling_2024}.

The benchmark simulation is carried out with square shearing sheets of increasing size.  There is no parallelisation involved in this benchmark. The kernel used for this benchmark is the \texttt{normal} kernel. This way, the potential computed by both solvers is the same. We initialise the domain with a single shearing wave like \cite{paardekooper_numerical_2012} and as described in the next section. With increasing problem size, the cell size remain constant as the size of the entire domain is increased with the number of cells.

\begin{figure}
    \centering
    \includegraphics[width=\linewidth]{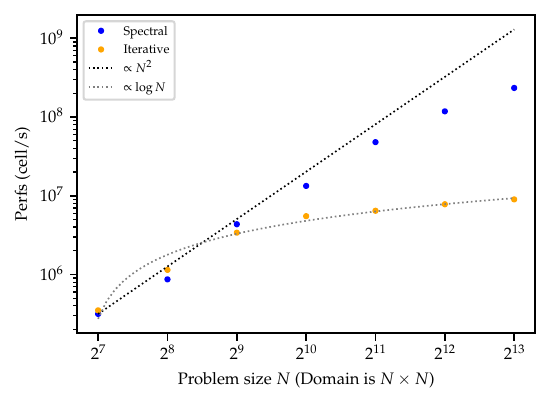}
    \caption{Performance comparison between the \texttt{PBICSTAB} solver implemented previously in \idefix and the spectral solver introduced in this paper. Higher is better.}
    \label{fig:Perfs}
\end{figure}

 We note that the last points of Fig. \ref{fig:Perfs} correspond to simulations that took too long and did not finish in less than 30 minutes. In all cases, the plotted performances is the average performance over the entire simulation.

As shown on Fig. \ref{fig:Perfs}, for larger problems, the spectral solver is much faster. From a problem size of $1024 \times 1024$ onward, we measure that the spectral solver is more than twice as fast as the iterative solver. The larger the number of cells, the faster the spectral solver gets compared to the iterative solver. In the largest case tested here ($8192\times 8192$) the spectral solver is 20 times faster than the Iterative solver.

Both solvers show sign of saturation at high domain size, but the spectral solver saturates both later and at higher level. The performance of the spectral solver scales almost linearly with the number of cells in the simulations ($\propto N^2$), whereas the iterative solver only scales logarithmically.

\section{Shear-periodic boundary condition test and resolution tests}
\label{app:BC}

Following \cite{paardekooper_numerical_2012}, we implement a shearing wave initial condition setup in \idefix to verify the implementation of the shear-periodic condition and self-gravity.

\begin{figure}
    \centering
    \includegraphics[width=0.9\columnwidth]{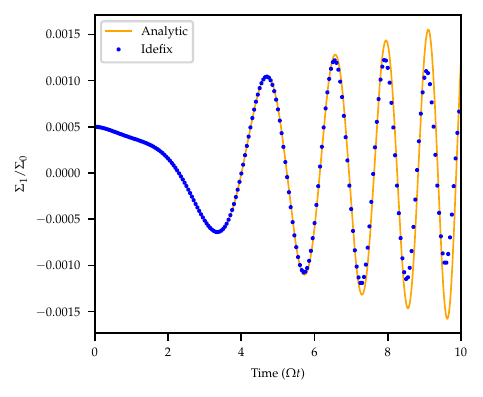}
    \caption{Shearing wave amplitude evolution \emph{without} self-gravity.}
    \label{fig:SWnoSG}
\end{figure}

\begin{figure}
    \centering
    \includegraphics[width=0.9\columnwidth]{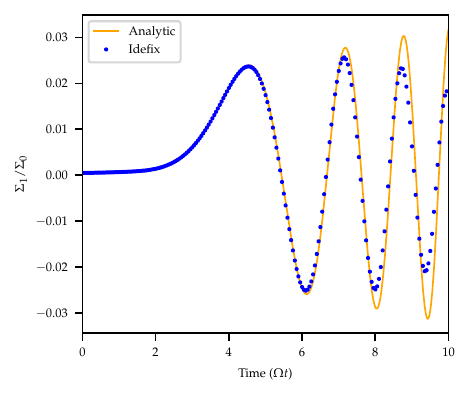}
    \caption{Shearing wave amplitude evolution \emph{with} self-gravity.}
    \label{fig:SWSG}
\end{figure}

Figures~\ref{fig:SWnoSG} and \ref{fig:SWSG} show the time evolution of the amplitude of an initial shearing wave perturbation, for the cases without and with self-gravity, respectively. Both figures compare the analytical solution (orange solid line) and the solution computed by \idefix (blue dots). The initial conditions are the same as in \cite{paardekooper_numerical_2012}, that is, 

\begin{equation}\;
	\begin{cases}
		\;\sigma(0) &= 0.0005\\
		\;\Sigma_0 & = 1/40\\
		\;k_{x,0} & = -4\pi \\
		\;k_y & = 2 \pi \\
	\end{cases}
\end{equation}

\begin{figure}
    \centering
    \includegraphics[width=0.9\columnwidth]{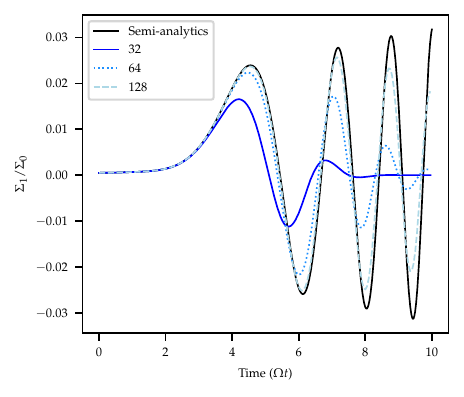}
    \caption{Convergence test with regards to numerical resolution. Our figure is analogous to Fig.~3 of \cite{paardekooper_numerical_2012}.}
    \label{fig:ResConv}
\end{figure}

Figures \ref{fig:SWnoSG} and \ref{fig:SWSG} have a resolution of $128\times128$ cells with box size $L_x = L_y = 1$ and where $H = \cs / \Omega \approx 0.07$, or approximately 9 points per scaleheight $H$. Figure \ref{fig:ResConv} uses the same domain size and $H$ but with also less grid points.

We also perform a resolution test, like \cite{paardekooper_numerical_2012} and show it in Figure~\ref{fig:ResConv}: The behaviour we observe is the same as \cite{paardekooper_numerical_2012}, where too low resolution smooths out the structures. Regarding the simulations presented in this work, low resolution is not an issue for two reasons. First we use 64 points per scaleheight, $H$, which is more than 6 times as much as in the 128 case presented here. Second, we are interested in scales much larger than $H$. Theses modes are very well resolved by our grid with up to 2048 grid cells per wave-length for the $m=1$ mode.

The good agreement of the the theoretical shearing-wave mode evolution and the evolution produce by \idefix confirms that the shear-periodic boundary condition and self-gravity solver behave as expected.

This test is now integrated in the continuous integration routines of the \idefix code and continuously tested using MPI parallelisation and GPU execution, and for both spectral and iterative solvers.

\section{Box size effect}

In this appendix, we explore the effect of the box size used for our simulations. We produce a smaller box simulation and a bigger box simulation, as listed in Tab. \ref{tab:BoxSize}.

\begin{table}[h!]
	\centering
	\caption{List of simulation for the box size effect study}
	\label{tab:BoxSize}
	\begin{tabular}{ccccl}
		\hline
		$\Sigma_0$ & $\Gamma$ & $L_x/H \times L_y/H$ & $N_x \times N_y$ & Name$\vphantom{\int}$\\
		\hline
		0.80 & 1.1 & $32 \times 32$ & $2048 \times 2048$  & fiducial (F)  \\
        0.80 & 1.1 & $64 \times 64$ & $4096 \times 4096$  & big box (BB)  \\
        0.80 & 1.1 & $16 \times 16$ & $1024 \times 1024$  & small box (SB) \\ 
		\hline
	\end{tabular}
\end{table}

\begin{figure}
    \centering
    \includegraphics[width=\columnwidth]{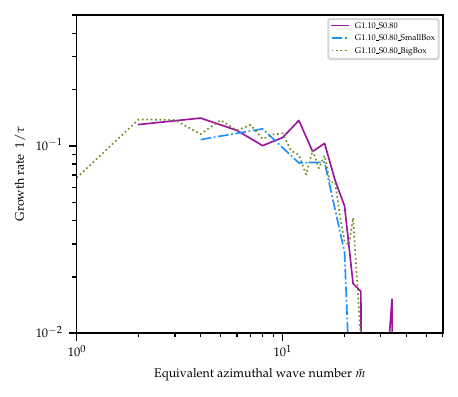}
    \caption{Growth rate of the azimuthal spiral modes, assuming $\vert \tilde{\Sigma}_m \vert \propto e^{t/\tau}$, as a function of their wave number for $\Omega t < 15$ for different box sizes. The mode number $\tilde{m}$ corresponds to that of the bigger box simulation. The other lines have been shifted so that the spacial wavelength coincide as discussed in the text body.}
    \label{fig:BoxSize}
\end{figure}

Figure \ref{fig:BoxSize} show how the spiral growth rate depending on the box size. On this figure, the azimuthal wave number $\tilde{m}$ is shifted such that they represent the same physical scales. This way, $\lambda_m^{\mathrm{F}} \neq \lambda_m^{\mathrm{SB}} \neq \lambda_m^{\mathrm{BB}}$, but $\lambda_{\tilde{m}}^{\mathrm{F}} = \lambda_{\tilde{m}}^{\mathrm{SB}} = \lambda_{\tilde{m}}^{\mathrm{BB}}$ the scaled wave-length are equal.
We see that the growth rates behave in the very same way for all simulations, which confirms that the box size plays little effect on the simulation.

We note, however, that the biggest box simulation show that its largest mode growth slower that for other simulation. This mode with wavelength $\lambda_0^{\mathrm{BB}}=64 H$ is only captured in this simulation. In this simulation it is not the $m=1$ mode that dominates, but the conclusions of this paper is not changed as it is still the largest scales modes that dominate.

We also note that extending the box size to 64 $H$ starts to reach the limit of the local approximation anyway. Indeed in this approximation, we neglect curvature terms, this holds if we study scales $L\ll \mathcal{P}$, where $\mathcal{P}$ is the perimeter of the disc at the radius of the Hill approximation $R_0$.
Quantitatively, we have $\mathcal{P}=2\pi R_0 = 2 \pi \varepsilon^{-1} H $, where $\varepsilon = H/R_0 = \cs/(\Omega R_0) $ is the hydrostatic aspect ratio of the disc. Proto-planetary discs are usually assumed to have constant aspect ratio $\varepsilon \gtrsim 0.05 - 0.1$, with larger values at larger radii \citep{andrews_ppd_2009,lesur_mhd_2021}. This means that $\mathcal{P}/H = 2\pi \varepsilon^{-1} \lesssim 63 - 126 $. This suggests that the largest scale modes with $\lambda_m > 32 H$ will be sensitive to global effects not captured here, as they would represent scales ranging from a quarter to half the disc perimeter at their radius. Hence, the change in behaviour for this specific mode for the biggest box simulation should be interpreted with caution.

\label{LastPage}

\end{document}